# More on Combinatorial Batch Codes

Sushmita Ruj and Bimal Roy
Applied Statistics Unit, Indian Statistical Institute
203 B T Road, Kolkata 700 108, INDIA
Email: {sush_r, bimal}@isical.ac.in

**Abstract**

Paterson, Stinson and Wei [2] introduced Combinatorial batch codes, which are combinatorial description of Batch code. Batch codes were first presented by Ishai, Kushilevita, Ostrovsky and Sahai [1] in STOC'04. In this paper we answer some of the questions put forward by Paterson, Stinson and Wei and give some results for the general case $t > 1$ which were not studied by the authors.

## 1 Introduction

Batch codes were proposed by Ishai, Kushilevita, Ostrovsky and Sahai [1]. Suppose there exists a large database of $n$ items distributed in $m$ devices. A user has to choose an arbitrary subset of $k$ items such that she can read at most $t$ items from each device. The problem is to distribute the items in the database such that the total storage $N$ is minimized. Batch codes are constructed for this purpose. Paterson, Stinson and Wei [2] gave a combinatorial formulation of the above problem and introduced the study of Combinatorial batch codes. They studied the special case when $t = 1$. In this paper we extend their work to study combinatorial batch codes for arbitrary $t$. We also give solutions to some of the questions raised by them.

*Batch codes* can be defined as follows:

**Definition 1** *[2, Section 1] An $(n, N, k, m, t)$ batch code over an alphabet $\sum$ encodes a string $x \in \sum^n$ into an $m-tuple$ of strings $y_1, y_2, \ldots, y_m \in \sum^*$ (also referred to as servers) of total length $N$, such that for each $k-tuple$ (batch) of distinct indices $i_1, i_2, \ldots, i_k \in \{1, 2, \ldots, n\}$, the entries $x_{i_1}, x_{i_2}, \cdots, x_{i_k}$ from $x$ can be decoded by reading at most $t$ symbols from each server.*

Our aim is to minimize $N$. In [2], Paterson, Stinson and Wei initiated the study of *Combinatorial Batch Codes* which is combinatorial formulation of the replication based Batch codes as proposed by Ishai, Kushilevita, Ostrovsky and Sahai [1]. Combinatorial Batch codes can be defined as follows:

**Definition 2** *[2, Section 1] An $(n, N, k, m, t)$ combinatorial batch code (CBC) is a set system $(X, \mathcal{B})$, where $X$ is a set of $n$ elements (called items), $\mathcal{B}$ is a collection of $m$ subsets of $X$ (called servers) and $N = \sum_{B \in \mathcal{B}} |B|$, such that for each $k$-subset $\{x_{i_1}, x_{i_2}, \ldots, x_{i_k}\} \subset X$ there exists a subset $C_i \subseteq B_i$, where $|C_i| \leq t$, $i = 1, 2, \ldots, m$, such that*

$$\{x_{i_1}, x_{i_2}, \cdots, x_{i_k}\} \subset \bigcup_{i=1}^{m} C_i$$

Combinatorial batch codes can be represented as follows : The items are represented by points of the set system and the servers are represented by subsets of these points. We follow the conventions as given in [2]. We consider the *dual set system*, in which the servers are represented by points and each item in a database is represented by a set (referred to as a *block*) containing the points (i.e., the servers) that stores that item.

Given a set system $(X, \mathcal{B})$ with $X = \{x_1, x_2, \cdots, x_n\}$ and $\mathcal{B} = (B_1, B_2, \cdots, B_b)$, the *incidence matrix* of $(X, \mathcal{B})$ is a $b \times n$ matrix $A = (a_{i,j})$, where

$$a_{i,j} = \begin{cases} 1 \text{ if } x_j \in B_i \\ 0 \text{ if } x_j \notin B_i \end{cases}$$

Paterson, Stinson and Wei considered the special case where $t = 1$. They represented $(n, N, k, m, 1)$ by $(n, N, k, m)$-CBC. We also follow this convention. Additionally, we represent $(n, N, k, m, t)$ by $(n, N, k, m)_t$-CBC for $t > 1$. The $(n, N, k, m)$-CBC is *optimal* if $N \leq N'$ for all $(n, N', k, m)$-CBC and we denote the corresponding value of $N$ by $N(n, k, m)$. Similarly we denote the optimal value of a $(n, N, k, m)_t$-CBC by $N_t(n, k, m)$ for $t > 1$.

A set if positions in a matric is called a *transversal*, if it contains one position in each row and in each column. We state the important lemma given in [2].

**Lemma 1** *[2][Section 1.1] An $m \times n$ 0-1 matrix is an incidence matrix of an $(n, N, k, m)$-CBC if and only if, for any $k$ columns, there is a $k \times k$ sub matrix which has at least one transversal containing $k$ ones.*

Refer to exmaple 1. If we consider any four columns say 12, 25, 40, and 55, then a transversal can be obtained as shown in the figure 1 below.

$$\begin{array}{cccc} 0 & 1 & 0 & 0 \\ \boxed{1} & 0 & 0 & 1 \\ 1 & \boxed{1} & 0 & 0 \\ 0 & 1 & 1 & \boxed{1} \\ 1 & 0 & \boxed{1} & 0 \\ 0 & 0 & 1 & 1 \end{array}$$

Figure 1: One way of choosing $k$ columns.

The rate of a code is defined as the ratio $c = n/N$. We define *column uniform Batch Codes* as batch codes which have the same number of ones in all the columns. If $c$ be the rate then there are exactly $1/c$ ones in each column. This also implies that each item is replicated $1/c$ times and stored in servers. We denote by $n(m, c, k)$ and $n_t(m, c, k)$, the maximum values of $n$ for which there exists a uniform $(n, cn, k, m)$-CBC and $n_t(n, cn, k, m)$-CBC respectively.

## 1.1 Previous results

We present the results given by Paterson, Stinson and Wei [2].

1. $N(n, k, n) = n$

2. $N(n, k, k) = kn - k(k-1)$

3. $N(n, k, n-1) = n - 1 + k$

4. If $n \geq (k-1)\binom{m}{k-1}$, then $N(n, k, m) = kn - (k-1)\binom{m}{k-1}$.

For fixed rate combinatorial batch codes, the authors stated the following results in [2].

1. $n(m, c, k) \leq \frac{k-1\binom{m}{c}}{\binom{k-1}{c}}$

2. $n(m, c, c+1) = c\binom{m}{c}$

3. $n(m, c, c+2) = \binom{m}{c}$

4. $n(m, 2, 4) = \binom{m}{2}$

5. $\lceil (m^2 - 1)/4 \rceil \leq n(m, 2, 5) \leq \lceil (m^2 + 2m - 3)/4 \rceil$.

## 1.2 Our Results

In [2] the authors asked the following questions. Can $N(n, k, m)$ be computed for a range of values of $n$, where $n < (k-1)\binom{m}{k-1}$? We answer this question by finding the optimal value $N(n, k, m)$ when $\binom{m}{k-2} \leq n \leq (k-1)\binom{m}{k-2}$. We also present some column uniform batch codes with fixed rate 1/3. This question was also posed by Paterson, Stinson and Wei.

In their paper [2], the authors considered the case $t = 1$. We present some result for arbitrary $t$.

## 2 Optimal Value $N(n, m, k)$ when $\binom{m}{k-2} \leq m \leq (k-1)\binom{m}{k-1}$

Paterson, Stinson and Wei gave some exact values of $N(n, m, k)$ for $n \geq (k-1)\binom{m}{k-1}$. They showed that if $n \geq (k-1)\binom{m}{k-1}$, then $N(n, k, m) = kn - (k-1)\binom{m}{k-1}$.

We extend this result for values of $n$ such that $\binom{m}{k-2} \leq n \leq (k-1)\binom{m}{k-2}$.

We say that $s$ columns span $l$ rows if the $s$ blocks span $l$ points. This means that $s$ blocks together contain $l$ points. Putting another way, $s$ items are contained in $l$ servers. First we try to find the exact values of $N(n, m, k)$ for $n$ close to $(k-1)\binom{m}{k-1}$. We observe that when $n = (k-1)\binom{m}{k-1}$, then each column of the incidence matrix contains $k-1$ ones and is repeated $k-1$ times. The reason behind this is that any two distinct columns span at least $k$ rows. If $k$ columns are selected at least two will be distinct and hence will span $k$ rows. We say that $\binom{m}{k-1}$ distinct columns belong to the same *group*. So there are $k-1$ groups.

We denote a column $i$ of the incidence matrix of a CBC by $(a_{1,i} a_{2,i} \ldots a_{m,i})^T$.

If we consider all the $\binom{m}{k-1}$ columns in a given block then, the number of columns which will have ones in the same $k-1$ rows is 1, the number of distinct columns which will have ones in the same $k-2$ rows will be $m - (k-2)$. The number of distinct columns which will have ones in the same $i$ rows will be $\binom{m-i}{k-i-1}$.

Consider the Example.

**Example 1** *Let $m = 6$, $k = 4$. Let $n = (k-1)\binom{m}{k-1}$. The 60 columns each contain three ones and three zeros.*

| 1 | 2 | 3 | 4 | 5 | 6 | 7 | 8 | 9 | 10 | 11 | 12 | 13 | 14 | 15 | 16 | 17 | 18 | 19 | 20 |
|---|---|---|---|---|---|---|---|---|----|----|----|----|----|----|----|----|----|----|----|
| 1 | 1 | 1 | 1 | 1 | 1 | 1 | 1 | 1 | 1  | 0  | 0  | 0  | 0  | 0  | 0  | 0  | 0  | 0  | 0  |
| 1 | 1 | 1 | 1 | 0 | 0 | 0 | 0 | 0 | 0  | 1  | 1  | 1  | 1  | 1  | 1  | 0  | 0  | 0  | 0  |
| 1 | 0 | 0 | 0 | 1 | 1 | 1 | 0 | 0 | 0  | 1  | 1  | 1  | 0  | 0  | 0  | 1  | 1  | 1  | 0  |
| 0 | 1 | 0 | 0 | 1 | 0 | 0 | 1 | 1 | 0  | 1  | 0  | 0  | 1  | 1  | 0  | 1  | 1  | 0  | 1  |
| 0 | 0 | 1 | 0 | 0 | 1 | 0 | 1 | 0 | 1  | 0  | 1  | 0  | 1  | 0  | 1  | 1  | 0  | 1  | 1  |
| 0 | 0 | 0 | 1 | 0 | 0 | 1 | 0 | 1 | 1  | 0  | 0  | 1  | 0  | 1  | 1  | 0  | 1  | 1  | 1  |

| 21 | 22 | 23 | 24 | 25 | 26 | 27 | 28 | 29 | 30 | 31 | 32 | 33 | 34 | 35 | 36 | 37 | 38 | 39 | 40 |
|----|----|----|----|----|----|----|----|----|----|----|----|----|----|----|----|----|----|----|----|
| 1  | 1  | 1  | 1  | 1  | 1  | 1  | 1  | 1  | 1  | 0  | 0  | 0  | 0  | 0  | 0  | 0  | 0  | 0  | 0  |
| 1  | 1  | 1  | 1  | 0  | 0  | 0  | 0  | 0  | 0  | 1  | 1  | 1  | 1  | 1  | 1  | 0  | 0  | 0  | 0  |
| 1  | 0  | 0  | 0  | 1  | 1  | 1  | 0  | 0  | 0  | 1  | 1  | 1  | 0  | 0  | 0  | 1  | 1  | 1  | 0  |
| 0  | 1  | 0  | 0  | 1  | 0  | 0  | 1  | 1  | 0  | 1  | 0  | 0  | 1  | 1  | 0  | 1  | 1  | 0  | 1  |
| 0  | 0  | 1  | 0  | 0  | 1  | 0  | 1  | 0  | 1  | 0  | 1  | 0  | 1  | 0  | 1  | 1  | 0  | 1  | 1  |
| 0  | 0  | 0  | 1  | 0  | 0  | 1  | 0  | 1  | 1  | 0  | 0  | 1  | 0  | 1  | 1  | 0  | 1  | 1  | 1  |

| 41 | 42 | 43 | 44 | 45 | 46 | 47 | 48 | 49 | 50 | 51 | 52 | 53 | 54 | 55 | 56 | 57 | 58 | 59 | 60 |
|----|----|----|----|----|----|----|----|----|----|----|----|----|----|----|----|----|----|----|----|
| 1  | 1  | 1  | 1  | 1  | 1  | 1  | 1  | 1  | 1  | 0  | 0  | 0  | 0  | 0  | 0  | 0  | 0  | 0  | 0  |
| 1  | 1  | 1  | 1  | 0  | 0  | 0  | 0  | 0  | 0  | 1  | 1  | 1  | 1  | 1  | 1  | 0  | 0  | 0  | 0  |
| 1  | 0  | 0  | 0  | 1  | 1  | 1  | 0  | 0  | 0  | 1  | 1  | 1  | 0  | 0  | 0  | 1  | 1  | 1  | 0  |
| 0  | 1  | 0  | 0  | 1  | 0  | 0  | 1  | 1  | 0  | 1  | 0  | 0  | 1  | 1  | 0  | 1  | 1  | 0  | 1  |
| 0  | 0  | 1  | 0  | 0  | 1  | 0  | 1  | 0  | 1  | 0  | 1  | 0  | 1  | 0  | 1  | 1  | 0  | 1  | 1  |
| 0  | 0  | 0  | 1  | 0  | 0  | 1  | 0  | 1  | 1  | 0  | 0  | 1  | 0  | 1  | 1  | 0  | 1  | 1  | 1  |

*If we consider all the 20 columns in a block then, the number of columns which will have ones in the same $k-1$ rows is 1. The number of columns which will have ones in the same $k-2 = 2$ rows will be $m - (k-2) = 4$. For example columns 10, 16, 19 and 20 have ones in the rows 5 and 6. No other column in the same group has ones in these two rows. The number of columns which will have ones in the same $k-3 = 1$ rows will be $\binom{m-(k-3)}{2} = 10$.*

While finding the values of $N(n, m, k)$ we decrease the value of $n$ by deleting columns at each step and modifying the existing columns such that any $k$ columns will span at least $k$ rows. We consider any column say $(\overbrace{00\cdots 0}^{m-k+1}\overbrace{11\cdots 1}^{k-1})^T$. We consider all columns in the same block which contain $k-2$ ones in the rows $m-k+3, m-k+4, \ldots, m$. There are $m-k+2$ such distinct columns. Suppose we delete less than $m-k+1$ of these distinct columns. Wlog, let the remaining columns be $(1\overbrace{00\cdots 0}^{m-k+1}\overbrace{11\cdots 1}^{k-2})^T$ and $(01\overbrace{00\cdots 0}^{m-k}\overbrace{11\cdots 1}^{k-2})^T$. Suppose we now change the first column to $(\overbrace{00\cdots 0}^{m-k+2}\overbrace{11\cdots 1}^{k-2})^T$. Then we find that if we select the column $(01\overbrace{00\cdots 0}^{m-k}\overbrace{11\cdots 1}^{k-2})^T$ $k-1$ times one each from the $k-1$ groups and the column $(\overbrace{00\cdots 0}^{m-k+2}\overbrace{11\cdots 1}^{k-2})^T$ once then the $k$ columns do not span $k$ rows. Hence we cannot change the first column. Similarly with the second column. If we delete less than $m-k+1$ columns, then we cannot change any of the other columns, because reducing the number of ones we can always find another column having $k-2$ ones in the same rows which when considered $k-1$ times from $k-1$ groups, together with the changed column will span less than $k$ rows.

This leads us to the following Theorem.

**Theorem 1** *If $n \geq (k-1)\binom{m}{k-1} - m + k$, then $N(n, m, k) = n(k-1)$.*

From the discussion above, the following observation can be made.

**Observation 1** If $m - k + 1$ columns in the same group are so deleted that they have $k - 2$ ones in the same rows, then the remaining column which contain $k - 2$ ones in the same row can be modified such that the $k - 2$ ones in those rows are unchanged and a zero is placed instead of 1 in the remaining position. We show that if such a construction is made, then any $k$ columns will span $k$ rows.

We consider the three possible cases.

1. If we consider any $k$ columns such that replica of the deleted/modified column(s) is not chosen, then any two of them will be distinct. As shown previously they will span $k$ rows.

2. If one or more columns are so selected that the replica of the deleted column has been selected and the modified column is not selected then atleast two distinct columns will be selected. So any $k$ columns will span at least $k$ rows.

3. So now we consider that the modified column is selected. Here two cases arise.

    (a) The $k - 2$ selected columns are the same as the modified column (chosen from the other groups) and one column is selected at random. Since there are atleast two distinct columns containing $k - 1$ ones, the $k$ columns will span $k$ rows.

    (b) The $k - 1$ columns are selected at random, then atleast two will be distinct. So the $k$ columns will span $k$ rows.

We note that there are $k - 2$ duplicate columns which will have ones in the same position as the modified column. These $k - 2$ columns together with the modified column will span $k - 1$ rows. However if we select any other row then all the $k$ columns will span $k$ rows. From this the Theorem follows.

**Theorem 2** *If $n = (k-1)\binom{m}{k-1} - (m - k + 1)$, then $N(n, m, k) = n(k-1) - 1$.*

We now present a construction which gives us the optimal value for any $n$ where $\binom{m}{k-2} \leq n \leq (k-1)\binom{m}{k-1}$. We then state the main theorem which gives us all values of $N(n, m, k)$ for $\binom{m}{k-2} \leq n \leq (k-1)\binom{m}{k-1}$.

We construction similar to that presented earlier in this section. We delete columns from the incidence matrix and see how we can reduce the number of ones in the existing columns.

**Construction 1** *Let the initial incidence matrix containing $(k-1)\binom{m}{k-1}$ columns be denoted by $A$.*

1. $m - k$ columns rea deleted from the same group $g_1$, such that all have ones in the $m - k + 3, m - k + 4, \cdots, m$ rows.

2. The columns $(\overbrace{00\cdots0}^{m-k+1}\overbrace{11\cdots1}^{k-1})^T$, $(\overbrace{00\cdots0}^{m-k}10\overbrace{11\cdots1}^{k-2})^T$, $\cdots$, $(01\overbrace{00\cdots0}^{m-k}\overbrace{11\cdots1}^{k-2})^T$ are deleted from the same group $g_1$ and the column $(1\overbrace{00\cdots0}^{m-k+1}\overbrace{11\cdots1}^{k-2})^T$ in group $g_1$ is modified to $(\overbrace{00\cdots0}^{m-k+2}\overbrace{11\cdots1}^{k-2})^T$.

3. Similarly, the columns $(\overbrace{00\cdots0}^{m-k}110\overbrace{11\cdots1}^{k-3})^T$, $(\overbrace{00\cdots0}^{m-k-1}1010\overbrace{11\cdots1}^{k-3})^T$, $\cdots$, $(01\overbrace{00\cdots0}^{m-k-1}10\overbrace{11\cdots1}^{k-3})^T$ are deleted from the same group $g_1$ and $(\overbrace{00\cdots0}^{m-k+1}\overbrace{11\cdots1}^{k-1})^T$ are deleted from any other group $g_2$ and the column $(1\overbrace{00\cdots0}^{m-k}10\overbrace{11\cdots1}^{k-3})^T$ in group $g_1$ modified to $(\overbrace{00\cdots0}^{m-k+1}10\overbrace{11\cdots1}^{k-3})^T$. So $m - k + 1$ columns are deleted at this step.

4. Similarly, we can delete the columns $(\overbrace{00\cdots0}^{m-k}1110\overbrace{11\cdots1}^{k-4})^T$, $(\overbrace{00\cdots0}^{m-k-1}10110\overbrace{11\cdots1}^{k-4})^T$, $\cdots$, $(01\overbrace{00\cdots0}^{m-k}0110\overbrace{11\cdots1}^{k-4})^T$ form the same group $g_1$ and $(\overbrace{00\cdots0}^{m-k+1}\overbrace{11\cdots1}^{k-1})^T$ from another group $g_3$ ( we note that the corresponding column in $g_2$ has already been deleted in the previous step) and modify the column $(1\overbrace{00\cdots0}^{m-k}110\overbrace{11\cdots1}^{k-4})^T$ in group $g_1$ to $(\overbrace{00\cdots0}^{m-k+1}110\overbrace{11\cdots1}^{k-4})^T$. So $m - k + 1$ columns are deleted at this step also.

5. Suppose we consider the columns having ones in the $m-k-i+2$ th and $m-k-i+3$-th row. We delete $m-k-i+1$ columns from the group $g_1$ and $i$ columns from other groups in a similar manner as above. We then modify the appropriate column.

6. At the end of $\binom{m-1}{k-2}$ steps we are left with $(k-2)\binom{m-1}{k-2}$ columns containing $k-1$ ones and $\binom{m-1}{k-2}$ columns containing $k-2$ ones.

7. We delete $(m-k+1)$ columns and modify one column in each step such that after $\binom{m-1}{k-3}$ steps we are left with $\binom{m}{k-2}$ columns each having $k-2$ ones.

We present an example to demonstrate the construction.

**Example1 contd.** If we delete less than $m - k + 1 = 3$ columns say columns 59,60, then $n = 58$ and $N = n(k - 1)$. We delete the columns 56,59,60 and modify the column 50, such that it now contains 2 ones in rows 5 and 6. So now $n = 57$ and $N = n(k - 1) - 1$, We delete the columns 58,55 and modify the column 49, such that it now contains 2 ones in rows 4 and 6. We see that the if the columns 20, 40, and modified columns 49 and 50 are selected, then the four columns span only three rows. By construction we delete column 40. So now $n = 54$ and $N = n(k - 1) - 2$. In the third step we delete the columns 57, 54 and 20 and modify the column 48 such that new column has only two ones in rows 4 an 5. For $n = 55, 56$, $N = n(k - 1) - 2$. This gives $n = 51$ and $N = n(k - 1) - 3$. In the next few steps we delete the following columns. We represent this with the help of a table 2.

**Lemma 2** *Construction 1 gives a CBC with optimal N for $\binom{m}{k-2} \leq n \leq (k-1)\binom{m}{k-1}$.*

| Step | Columns deleted | Column Modified | Column after modification | $n$ | $N$ |
|------|-----------------|-----------------|---------------------------|-----|-----|
| 1 | 56,59,60 | 50 | $(000011)^T$ | 57 | $n(k-1)-1$ |
| 2 | 58,55,40 | 49 | $(000101)^T$ | 54 | $n(k-1)-4$ |
| 3 | 57,54,20 | 48 | $(000110)^T$ | 51 | $n(k-1)-4$ |
| 4 | 53,39,38 | 47 | $(001001)^T$ | 48 | $n(k-1)-4$ |
| 5 | 52,19,37 | 46 | $(001010)^T$ | 45 | $n(k-1)-5$ |
| 6 | 51,18,17 | 45 | $(001100)^T$ | 42 | $n(k-1)-6$ |
| 7 | 36,35,33 | 44 | $(010001)^T$ | 39 | $n(k-1)-7$ |
| 8 | 16,34,32 | 43 | $(010010)^T$ | 36 | $n(k-1)-8$ |
| 9 | 15,14,31 | 42 | $(010100)^T$ | 33 | $n(k-1)-9$ |
| 10 | 13,12,11 | 41 | $(011000)^T$ | 30 | $n(k-1)-10$ |
| 11 | 1,2,3 | 4 | $(110000)^T$ | 27 | $n(k-1)-11$ |
| 12 | 5,6,21 | 7 | $(101000)^T$ | 24 | $n(k-1)-12$ |
| 13 | 8,22,25 | 9 | $(100100)^T$ | 21 | $n(k-1)-13$ |
| 14 | 23,26,28 | 10 | $(100010)^T$ | 18 | $n(k-1)-14$ |
| 15 | 24,27,29 | 30 | $(100001)^T$ | 15 | $n(k-1)-15$ |

Table 1: Steps of Construction 1

**Proof :** We first show that the construction gives a valid $CBC$ and then show that the construction gives the optimal value of $N$ for any $n$ in the given range.

From the Theorem 2.9 and Lemma 3.5 given in [2] we find that that the optimal values for $n = (k-1)\binom{m}{k-1}$ and $n = \binom{m}{k-2}$ hold. Steps 1 and 2 have been shown to result in a CBC with optimal $N$ (Refer to Theorem 1 and Theorem 2. The reason for selecting $m-k$ columns from the same group and one column from another group for deletion in Step 3 is that, if we choose the column $(\overbrace{00\cdots 0}^{m-k+1}10\overbrace{11\cdots 1}^{k-3})^T$, $(\overbrace{00\cdots 0}^{m-k+2}\overbrace{11\cdots 1}^{k-2})^T$, and the column $(\overbrace{00\cdots 0}^{m-k+1}\overbrace{11\cdots 1}^{k-1})^T$ $k-2$ times then the $k$ columns will span only $k-1$ rows, which is undesirable. So we have delete another column $(\overbrace{00\cdots 0}^{m-k+1}\overbrace{11\cdots 1}^{k-1})^T$ from one of the $k-2$ groups ( say group $g_2$). For a similar reason we delete $m-k$ columns from the same group and one of the $(\overbrace{00\cdots 0}^{m-k+1}\overbrace{11\cdots 1}^{k-1})^T$ columns in Step 4.

We see that we cannot further modify any column at each step in a manner similar to the Observation 1 following Theorem 1

We have already seen that there are $m-k+2$ columns within the same group which have ones in the same $k-2$ rows. When we consider the columns which have a one in the $m-k+2$-th row in the same group then we note that $i$ such columns in the same group have been already deleted in the previous steps. So we delete $m-k+1-i = m-k-i+1$ of the remaining columns having one in the $m-k-i+2$-th row and $k-2$ ones in the same $k-2$ rows. We modify the remaining one column. However in this process we can observe that $k$ columns can be selected such that they span only $k-1$ rows. Hence $i$ other columns have to be deleted from other groups ( as done in step 4). Thus each step we delete $m-k+1$ columns.

After $\binom{m-1}{k-2}$ steps we will be left with $(k-1)\binom{m-1}{k-2}$, such that $\binom{m-1}{k-2}$ columns will have $k-2$ ones and the rest have $k-1$ ones (with one in the first column). Now we start deleting and modifying columns from the $(k-2)\binom{m-1}{k-2}$ columns such that at each step $m-k+1$ columns are deleted and one column is modified. So after $\binom{m-1}{k-3}$ steps we are left with $\binom{m-1}{k-3}$ new columns will have been modified. Thus there will now be $\binom{m-1}{k-2}+\binom{m-1}{k-2} = \binom{m}{k-2}$ columns, each having $k-2$ ones.

At each step the we get a CBC with optimal value of $N$ given $n$. So the entire construction gives a CBC with optimal $N$. ∎

From the above Lemma we arrive at the following theorem.

**Theorem 3** *If $\binom{m}{k-2} \leq n \leq (k-1)\binom{m}{k-1}$, $d = (k-1)\binom{m}{k-1} - n$ and $l = \lfloor \frac{d}{m-k+1} \rfloor$, then $N(n,m,k) = (k-1)n - l$.*

We next discuss about column uniform batch codes of rate 1/3.

## 3 Column Uniform Batch Codes with rate $1/3$

Column Uniform Batch codes with rate 1/3 are CBCs in which every block of the dual set system contains precisely 3 points. This means that every item is stored in exactly three servers.

We think of the points as vertices of a graph. Any three points which lie in the same block correspond to a triangle in the graph. For any block $b$ such that $a_{x,b} = a_{y,b} = a_{z,b} = 1$ and $a_{i,b} = 0$ for $i = \{1, 2, \ldots, m\}\setminus\{x, y, z\}$, there is a triangle $xyz$ in the corresponding graph. From Corollary 3.3 and 3.5 of [2] we see that

1. $n(m, 3, 4) = 3\binom{m}{3}$
2. $n(m, 3, 5) = \binom{m}{3}$

We next try to find some values for $n(m, 3, 6)$. The question of finding a valid $(n, 3n, k, m)$-CBC thus boils down to finding a graph such that any $k$ triangles in the graph must span atleast $k$ vertices. More importantly, we are interested in finding the graph on $n$ vertices which has the maximum number of triangles such that any $k$ triangles in the graph will span atleast $k$ points.

We give a construction for a column uniform CBC which has rate equal to 1/3. We note that the graph $K_4$ has four triangles. So any four triangles span atleast four vertices. Let $m = 8$. Then the graph $G_8$ (Refer to Figure 2) has 16 triangles such that any 6 triangles spans atleast 6 points. So there exists a $(16, 48, 6, 8) - CBC$.

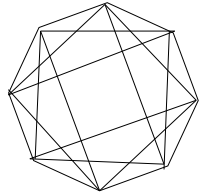

Figure 2: The graph $G_8$.

Now we consider the graphs $G_8$ and $K_4$ as units. The units have be interconnected as shown in the Figure 3.

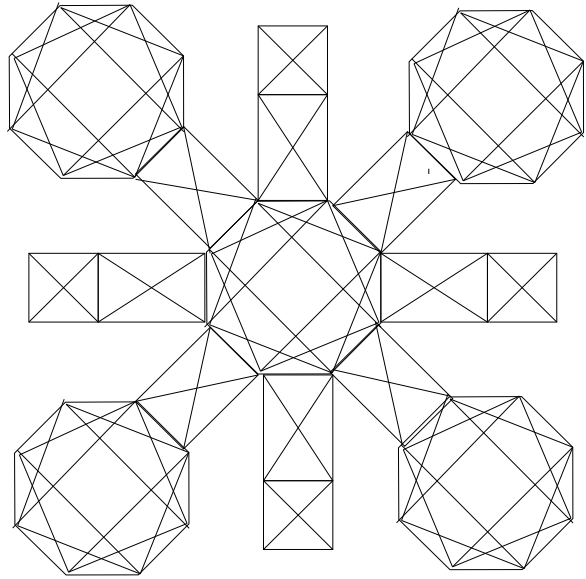

Figure 3: A unit consisting of $G_8$ and $K_4$

We replace the graphs $G_8$ and $K_4$ by an octagons and squares. If there are $g$ octagons and $s$ squares, then there will be $8g + 4s$ vertices and $16g + 4s + 4e$ triangles, where $e$ is the number of common edges between faces. The calculation of $e$ will be done in the following way.

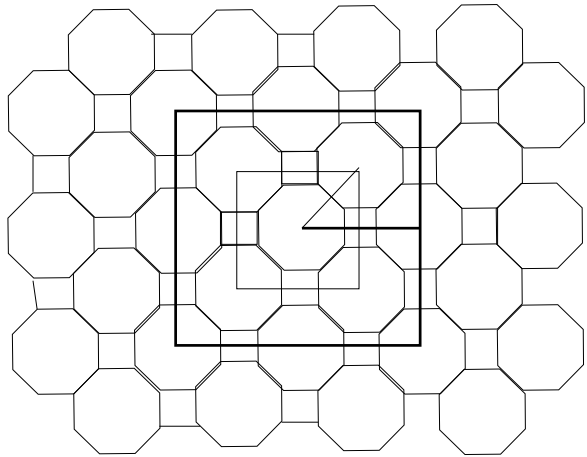

Figure 4: The pattern of octagon and square.

We refer to the Figure 4 below. Level $d$ denotes the octagons and squares at a distance $d$ from the central octagon. Consider a graph with only one level. We see that the number of octagons and squares at distance one is four each. Number of common edges between faces is 16. We thus note that $g = 5$, $s = 4$ and $e = 16$. So if $m = 8.5 + 4.4 = 56$, then the number of triangles present in the original graph is 160. So $n(56, 3, 6) = 16.5 + 4.4 + 64 \geq 160$. For a graph with level two, the number of octagons and squares is 8 each. So $g = 13$, $s = 12$, $e = 56$. So $(152, 3, 6) \geq 480$. To generalize at distance $d$, number of octagons and squares is $4d$ each.

$g = 1 + 2d(d + 1)$, $s = 2d(d + 1)$, $e = (8(g - 4d) + 4(s - 4d))/2 + 8d + (4d + 3(4d - 4) + 4)/2$.
Let $m = 8g + 4s$, then $n(m, 3, 6) \geq 16g + 4s + 4e$, where $g = 1 + 2d(d + 1)$, $s = 2d(d + 1)$, $e = 4g + 2s - 8d - 4$.

Given the value of $m$, we find lower bound for $(m, 3, 6)$ in the following way.

The distance can be calculated as $d = \lfloor \frac{24 + \sqrt{24^2 + 96(m-8)}}{48} \rfloor$. $g = 1 + 2d(d+1)$, $s = 2d(d+1)$, $e = 4g + 2s - 8d - 4$.

Let $\Delta' = 16g + 4s + 4e$, $m' = m - (8g + 4s)$, $j = m'(\text{mod } 8)$, then,

$$\Delta = \Delta' + oct(j) + l'(j) + \Delta''$$

where

$$oct(j) = \begin{cases} 0 \text{ for } j = 1, 2, \\ 1 \text{ for } j = 3, \\ 4 \text{ for } j = 4, \\ 5 \text{ for } j = 5, \\ 6 \text{ for } j = 6, \\ 10 \text{ for } j = 7, \\ 16 \text{ for } j = 8, \end{cases} \quad , l'(j) = \begin{cases} 1 \text{ for } j = 1 \\ 4 \text{ otherwise} \end{cases} \quad , \Delta'' = \max_{\substack{i=1 \\ to \lceil m'/8 \rceil}} \{8s' + 20i + 4l\},$$

where $s' = \lfloor \frac{m' - 8i}{4} \rfloor$, $l = \begin{cases} 7 \text{ for } j = 1, 2 \text{ and } i + s' + 1 = 8, \\ u + s' + i \text{ for } j \neq 1, 2 \text{ and } i + u + s' = 8, \\ u + s' + i - 1 \text{ otherwise.} \end{cases}$ and $u = \begin{cases} 0 \text{ for } j = 1, 2, \\ 1 \text{ otherwise} \end{cases}$.

## 4 Optimal values of $N$ for $t > 2$

We generalize the problem to find $N_t(n, m, k)$ which was not done in [2].

| 1 | 1 | 0 | 0 | 0 | 1 |
|---|---|---|---|---|---|
| 0 | 0 | 1 | 0 | 0 | 1 |
| 0 | 0 | 0 | 1 | 1 | 0 |

From the above matrix we see that we can four elements from any four columns by selecting a maximum of two elements from each row.

We give optimal value when $m = n$. This is a trivial case and for any $t$ we can state the following theorem.

**Theorem 4** $N_t(n, k, n) = n$

We next show the present the following example.

**Example 2** *The following represents the incidence matrix corresponding to $N_2(10, 5, 5)$.*

| 1 | 0 | 0 | 0 | 0 | 1 | 0 | 0 | 0 | 0 |
|---|---|---|---|---|---|---|---|---|---|
| 0 | 1 | 0 | 0 | 0 | 0 | 1 | 0 | 0 | 0 |
| 0 | 0 | 1 | 0 | 0 | 0 | 0 | 1 | 0 | 0 |
| 0 | 0 | 0 | 1 | 0 | 0 | 0 | 0 | 1 | 0 |
| 0 | 0 | 0 | 0 | 1 | 0 | 0 | 0 | 0 | 1 |

From the example it is very easy to see that

**Theorem 5** *If $n \leq tk$, then $N_t(n, k, k) = n$.*

Next we present results for column uniform batch codes with fixed rates. Let us consider the fixed rate $c = 2$. Suppose we can choose a maximum of $t$ elements from each server. We construct the incidence matrix such that each of the $\binom{m}{2}$ combinations is repeated. We can see that on choosing $2t$ columns in the worst case we can choose the same combination $2t$ times. Suppose this combination takes into consideration the first and the second rows. Then we can select $t$ ones each from the first and second rows. This essentially means we have chosen $2t$ items taking $t$ item from each of the first and second servers. From this we arrive at the following theorem.

**Theorem 6** $n_t(m, 2, 2t) = \infty$.

As a consequence for $k < 2t$, we can choose the $k$ items in any way we like such that at most $t$ items are selected from each of the servers(columns).

**Corollary 1** *If $k \leq 2t$, $n_t(m, 2, k) = \infty$.*

**Theorem 7**    1. $n_t(m, c, ct + i) = ct\binom{m}{c}$, for $i \leq t$

2. $n_t(m, c, ct + t + j) = t\binom{m}{c}$, for $j \leq t$

3. $n_t(m, c, k) < \binom{m}{c}$, if $k > tm$

**Proof :** There are $\binom{m}{c}$ distinct columns each having $c$ ones. We show that $n_t(m, c, ct + i) \geq ct\binom{m}{c}$, for $i \leq t$. We can replicate each of the $\binom{m}{c}$ columns $ct$ times. Now if we select any $ct$ columns columns which are replicas then we can choose $t$ ones from each of the $c$ rows and get a valid CBC. So $n_t(m, c, ct) \geq ct\binom{m}{c}$. If $k = ct + i$, $i \leq t$, then we can chosen $ct$ columns which are replicates and any other $i$ columns. Suppose we choose the column $(\overbrace{11\cdots 1}^{c}\overbrace{00\cdots 0}^{m-c})^T$ $ct$ times and the column $(\overbrace{11\cdots 1}^{c-1}01\overbrace{00\cdots 0}^{m-c-1})^T$ $i$ ($i \leq t$ times then the ones can be selected from the $c + 1$ rows a maximum of $t$ times. So we get a valid CBC. Any other choice of $ct + i$ columns will also give rise to a valid CBC. So $n_t(m, c, ct + i) \geq ct\binom{m}{c}$, when $i \leq t$.

We show that $n_t(m, c, ct + i) \leq ct\binom{m}{c}$, when $i \leq t$. Suppose atleast one of the $\binom{m}{c}$ columns be repeated $ct + 1$ times. For $k > ct$, let the choose columns include all the $ct + 1$ columns. Then at least one server contributes more than $t$ items, which is a contradiction. So $n_t(m, c, ct + i) \leq ct\binom{m}{c}$, when $i \leq t$.

Hence the Part 1 of the theorem follows.

The proof of Part 2 is similar to Part 1.

Consider $k > tm$. If $n_t(m, c, k) \geq \binom{m}{c}$, then any $k$ columns will span only $m$ rows. Since $k > tm$, at least any two rows contribute more than $t$ elements, which violates the condition that of combinatorial batch codes. ∎

Corollary 3.3 and 3.5 of [2] is a special case of the above Theorem 7 when $t = 1$.

**Comment** We give below a table which compares the storage $N$ for $t = 1$ and $t = 2$. We observe that the storage space required when $t = 1$ is about two times than when $t = 2$. This shows that for storage efficiency we can increase the number of probes. However we may note that for load balancing the number of probes cannot be too large.

| $n$ | $m$ | $k$ | $N(t=1)$ | $N(t=2)$ |
|-----|-----|-----|----------|----------|
| 180 | 10  | 5   | 640      | 360      |
| 180 | 10  | 6   | 684      | 360      |
| 720 | 10  | 7   | 4185     | 2160     |
| 240 | 10  | 9   | 1860     | 720      |

From Theorem 7 it follows that

1. $n_2(m,2,5) = 4\binom{m}{2}$
2. $n_2(m,2,6) = 4\binom{m}{2}$
3. $n_2(m,2,7) = 2\binom{m}{2}$
4. $n_2(m,2,8) = 2\binom{m}{2}$

We next try to find out the values of $n_2(m,2,9)$ and $n_2(m,2,10)$. We note that $n_2(m,2,k) \leq \binom{m}{2}$, $k > 10$.

We look at the graph in Figures 5(a) and 5(b). The graph does not have any triangle or square.

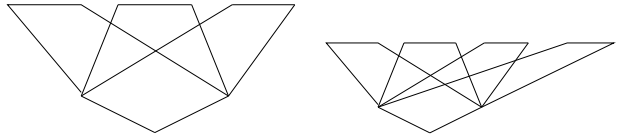

(a) Square and triangle free graph on 9 vertices   (b) Square and triangle free graph on 11 vertices

Figure 5: Square triangle free graphs

From this the following Lemma 3 follows. The lemma is important in proving Theorem 8.

**Lemma 3** *Let $G$ be a graph on $v$ vertices. Then the maximum number of edges such that $G$ does not any triangle or square is greater than $2 + 3\lfloor (v-3)/2 \rfloor$.*

**Theorem 8**  *1. $n_2(m,2,9) \geq \binom{m}{2} + 2\lfloor m/3 \rfloor$.*

*2. $n_2(m,2,10) \geq \binom{m}{2} + 2 + 3\lfloor (m-3)/2 \rfloor$.*

**Proof :** If we consider all possible two combinations, then there are $\binom{m}{2}$ columns. We note that $\binom{5}{2} = 10$. So if 9 elements are selected atleast five servers must contribute maximum two elements. So $n_2(m,2,9) \geq \binom{m}{2}$. Similarly for $n_2(m,2,10) \geq \binom{m}{2}$.

Let the incidence matrix containing all the $\binom{m}{2}$ columns be denoted by $A$. Now we try to find if we can add more columns containing two ones each. We represent the new augmented incidence matrix by $A' = (a'_{i,j})$. We map the incidence matrix to a graph, such that columns represent the edges. An edge $i$ is incident on two vertices $e$ and $f$ if $a'_{e,i} = a'_{f,i} = 1$. We have considered the cases when all the $k$ items represent columns in $A$. Now we consider some items to be selected from $A'$ as well. $A'$ is a sub matrix of $A$ constructed taking some columns of $A$.

We first prove Part 2. We are interested in finding the maximum number of columns that we must select from $A$ to form $A'$ such that all the conditions of combinatorial batch codes are met. If all the $k$ elements are selected from $A'$, then we have a valid code. So we select some elements from $A$ and some from $A'$. In particular we consider the condition where the same row is selected from the matrices $A$ and $A'$. The length of $A'$ can be analyzed once we consider this situation. We note that if we select items such that they belong to five or more servers, then it is clear that a maximum of two items are selected from each server. So we consider the situation in which items are selected from four servers. Now we try to see which columns can be replicated. At the maximum six columns can be selected from $A$ such that the columns span only four rows. If another four columns are so chosen as shown that they belong to the same row, then the ten chosen columns will span at the maximum four rows giving an invalid CBC. We call such combinations the *forbidden configurations*. Thus columns can be replicated such that the underlying graph does not contain any subgraph as the forbidden configurations. The forbidden configurations are graphs on four vertices with four edges as shown in Figure 6.

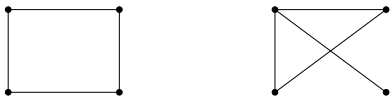

Figure 6: Forbidden configuration : graphs on four vertices which contain four edges.

From the graphs we see that there can be no triangle or square. From Lemma 3 we see that the maximum number of edges in a graph having no triangle or square is $2+3\lfloor(v-3)/2\rfloor$. Thus $n_2(m,2,10) \geq \binom{m}{2} + 2 + 3\lfloor(m-3)/2\rfloor$.

While finding a lower bound for $n(m,2,9)$ we see that the following are the forbidden configurations (Refer to Figure 7). We find that the graphs which do not have subgraphs as any of the configurations. Such graphs will consist of disjoint paths of length two. The number of edges on $m$ vertices is $2\lfloor m/3 \rfloor$. So $n(m,2,9) \geq \binom{m}{2} + 2\lfloor m/3\rfloor$. ∎

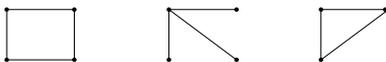

Figure 7: Forbidden configuration : graphs on four vertices which contain three edges.

## 5 Conclusion and open problems

In this paper we answer some of the questions put forward by Paterson, Stinson and Wei in [2]. We also study $N(n,m,k)$ for $t>1$. The following questions remain unsolved.

1. What is the optimal value of $N(n,m,k)$ for $n < \binom{m}{k-2}$.

2. How close are the bounds for codes $(n, 3n, m, k)$-CBC.

3. Finding optimal solutions for $N_t(n,m,k)$.